# Topographic maps in the brain are fundamental to processing of causality

Shigeko Takahashi
Kyoto City University of Arts

**Abstract**

The ubiquity of topographic maps in the brain has long been known, and molecular mechanisms for the formation of topographic organization of neural systems have been revealed. Less attention has been given to the question of 'Why are the maps topographical and why so ubiquitous?' In this study, I explore the implications of the topographic maps for brain function, by employing the mathematical framework of the Zeeman topology. I propose the notion about the meaning of topographic order as generic mechanisms for the representation and analysis of causal structure, implemented by the neural systems. This leads to a much improved understanding of the division of labour between chemical systems and neural systems in the formation of maps to deal with causality.

## 1. Introduction

Neural systems are well-ordered, non-random structural organization, and neural circuits display highly ordered projections between different brain areas. Many axonal connections in the brain, in particularly sensory systems, are topographically organized such that neighborhood relations are preserved from one neuronal population to another. General principles of the topographic organization in the brain have been revealed and the Eph tyrosine kinase receptors and their membrane-bound ligands (the ephrins ) are best known in their role in axon guidance, suggesting conserved mechanisms in the development of topographic maps in vertebrate and invertebrate nervous systems. It has been shown that the graded expression of these proteins is instrumental in providing molecular coordinates that define topographic maps, particularly in the visual system, but also in the auditory, vomeronasal and somatosensory systems as well as in the hippocampus, cerebellum and other structures (Palmer and Klein, 2003). These topographic maps require instructive signals from ephrins and Ephs, although the mode of action of this signaling family appears different between maps.

In the formation of visual maps, known as retinotopy, two subfamilies, the EphB/Efnbs and the EphA/Efnas, are expressed in complementary gradients between interconnected structures (i.e., retina and the superior colliculus (SC)/tectum, or V1 and SC/tectum) and as reciprocal gradients within structures (i.e., retina). Thus, retinal ganglion cells send their axons according to the molecular coordinates so that the spatial organization in the retina is maintained on SC and the visual cortex (Savier, 2016, Savier and Reber, 2018). Topographic mapping has also been demonstrated in the limbic circuits, a system critical for learning, memory and emotional behaviors. Hippocampal neurons project topographically to their major subcortical projection target, the lateral septum. The rodent hippocampal mossy fibers exhibit prominent structural plasticity according to certain topography principles in a process requiring the EphA receptors tyrosine kinase (Klein, 2009, 2010; Galimberti et al., 2010).

Why are the maps topographical and why so ubiquitous? The topographical features of cortical and subcortical maps may be essential to much brain function implemented by the neural systems. This study puts forward the notion about the meaning of topographic order as generic mechanisms for the representation and analysis of causal structure of the world, implemented by the neural systems. This leads to a much improved understanding of distinctive features of the neural functions in the brain.

## 2. Causality

Causality plays a central role in the way which organisms structure the world where they live. The essence of any kind of adaptive and/or predictive behaviors is to capture causality, i.e., causal structure of the physical world. The most primitive mathematical structure associated with causality is a partial order. Zeeman was the first to give a possible solution to the problem of a causal topology on the physical world, i.e. Minkowski space, the real 4-dimensional space-time continuum of special relativity. Zeeman (1964) proved that causality is represented by a partial ordering on Minkowski space and that the group of all automorphisms preserving this partial ordering is generated by the inhomogeneous Lorentz group and dilations. He remarked that if we interpret the principle of causality mathematically as the set M (Minkowski space) together with the partial ordering, then the inhomogeneous Lorentz group appears naturally (with dilations and space reversal) as the symmetry group of M, and that therefore the basic invariance of physics, which are the representations of the inhomogeneous Lorentz group, follows naturally from the single principle of causality. Zeeman (1966) proposed that the fine topology (causal topology) on 4-dimensional Minkowski space, which is defined to be finest topology (with the most open sets), induces the 3-dimensional Euclidean topology on every space axis and the 1-dimensional Euclidian topology on every time axis. Note that the physical significance of the causal order is that one can propagate "*information*" from x to y if and only if $x \leqslant y$ ($\leqslant$ is a partial order). The physical mechanisms used to propagate "*information*" involve sending material particles and light signals.

On the basis of a mathematical theory of computation, Martin and Panangaden (2004, 2006) proved that a globally hyperbolic spacetime with its causality relation is a bi-continuous poset (a partially ordered set) whose interval topology is the manifold, and showed that from a countable dense discrete set equipped with a partial order relation interpreted as causality, one can reconstruct the spacetime manifold with its causal topology.

These arguments implies that causality is not emergent but a partial order set allows for the derivation of several intrinsically defined topology including causal topology of the physical world; From a dense discrete set with a partial order relation, spacetime can be order-theoretically reconstructed; the causal structure of the physical world emanates from something discrete, i.e., a partially ordered discrete set.

## 3. Topographic maps as a partially ordered set to represent causality

The topographically organized neural connections in the brain maps provide the mechanism whose function is mathematically defined as a poset (a dense discrete set with a partial order relation) by which causality is represented and/or processed in a computable manner.

In the visual system, the physical space (the visual field) is conformally mapped on the 2-dimensional layer of photoreceptors in the retina, and the retinal ganglion cells project in a topographic manner to several targets, mainly the superior colliculus (tectum in birds) in the midbrain and the lateral geniculate nucleus in the forebrain such that the retinotopic maps (i.e., topology) are maintained along the visual pathways to the visual cortex (area V1, V2,.and V3) (Savier, 2016; Savier and Reber, 2018). In the formation of the retinotopic maps, the graded expression of ephrins guide growing axons to distinct locations in a topographic projection: The projection of retinal axons is topographically ordered along two axes; depending on the location of a given neuron on dorso-ventral and naso-temporal axes of the retina, its axon will terminate at distinct positions along the medio-lateral and antero-posterior axes of the superior colliculus and the lateral geniculate nucleus, respectively. Individual retinal ganglion cells express different levels of type A Eph receptors along the naso-temporal axis, creating a smooth gradient of receptor expression and type A ephrin sensitivity (Frisen,et al.,1999). In addition to the expression of Eph receptors, several ligands are expressed in the retina, and some of them in a gradient. Thus, it is suggested that the responsiveness of a neuron on an ephrin is regulated not only by the level of receptor expression but also by co-expression of ligand. The expression patterns of ephrins and Eph receptors in the thalamo-cortical projections, septo-hippocampal tract, nigro-striatal pathway and motor neuron projections to muscles along the anterior-posterior axis implicate that these molecules are involved in the organization of the topography of these systems (Frisen et al., 1999; Donoghue et al., 1996; Gao et al., 1996; Yue et al., 1999, 2002).

The fact that the topographical features of the brain maps are defined by a smooth gradient of receptor expression and ligand sensitivity indicates that the population of neurons recruited into a map can function as a poset, because individual neurons receive distinct inputs from distinct positions in a topographic projection and the set of neurons is equipped with a partial ordering. Thus, the topographically organized neural maps provide the generic mechanisms to seek causal structure for the sensed physical world. This is possible because neighborhood relations (thus, topology) are preserved from one neuronal population to another along respective neural pathway, according to molecular coordinates. The division of labour between chemical systems and neural systems in the formation of maps plays a major role to deal with causality.

Moreover, the topography of the auditory map and its alignment with the visual map develops gradually, depending on the appropriate sensory experience during development (Feldman et al.,1996; Schnupp et al., 1995). The visual input dictates the alignment of the auditory and visual maps: N-methyl-D-aspartate (NMDA) receptor is important in the process of alignment and realignment (Rees, 1996). Thus, the topographical organization may not be simply computationally efficient, but essential to capture the causal structure of the world.

## 4. Topographic maps and hierarchical representation commensurability

The shared properties of topographic maps in the brain, that is, the formation of topographic maps is precisely regulated by smooth gradients of the level of Eph receptor expression and co-expression of ligand and a set of neurons in a map functions as a poset equipped with a partial ordering defined by the molecular coordinates, may solve the problem of the hierarchical representation commensurability: how heterogeneous hierarchical representations can be unified in the brain?

A well established view that the brain is divided into stratified levels of organization, where the representations are hierarchical and the properties of higher levels influence or make a difference on lower levels, leads us to seek a comprehensive account of the hierarchical representations. The situation is compounded by the aspects of hierarchical representations. One aspect involves the diversity of representations: hierarchical representations can pertain to compositional (spatial), functional (procedural) properties, and/or abstract properties (see Takahashi and Ejima, 2016). Another aspect is temporality: how temporal relations are designated, whether via absolute chronology or via some type of staging or periodization. We have thus far observed that the representation articulated for the sensory system such as the visual system is not commensurate with the representations involved in other sensory system of different modality or the memory system.

Most of the goals of the brain function are comprised in understanding and mastering the causal hierarchy of our environment, i.e., space-time. Therefore, the hierarchical representations in topographic maps in the brain should share increasingly better causal properties, each level with some specific results, and the levels should be developed from the lowest (modality specific) to the highest one (global, amodal and/or abstract). The topographical organization of the neural systems, precisely regulated via chemical signals, may provide a fundamental architecture to ensure the hierarchical representation commensurability for processing of causality.